\documentclass{aa}
\usepackage{graphicx}
\usepackage{txfonts}

\begin{document}
%
   \title{ Spectral properties of nonthermal X-ray emission from the
     shell-type SNR RX~J1713.7--3946 as revealed by XMM-Newton}


   \author{Junko S. Hiraga,
          \inst{1}
          Yasunobu Uchiyama,
         \inst{2}
         Tadayuki Takahashi
         \inst{1} 
          \and
          Felix A. Aharonian
          \inst{3}
          }

   \offprints{J.Hiraga,\\ \email{jhiraga@astro.isas.jaxa.jp}}

   \institute{Institute of Space and Astronautical Science, JAXA,  3-1-1 Yoshinodai, Sagamihara,
   Kanagawa 229-8510, Japan\ \
	\and
	Yale Center for Astronomy \& Astrophysics, P.O. Box 208121, New Haven, 
CT 06520-8121, USA\\	
         \and
             Max-Plank-Institut f\"ur Kernphysik, Postfach 103980, D-69026 Heidelberg, Germany\\
             }

   \date{submitted 24 December 2003; accepted 13 July 2004}

   \abstract{ We present the results of our morphological and spectral
   study of properties of the supernova remnant RX~J1713.7$-$3946
   based on data obtained with \emph{XMM-Newton}. Highly inhomogeneous
   structures, such as the bright spots, filaments and dark voids,
   noted by Uchiyama et al.~(2003), appear in the entire bright
   western portion of the shell. In addition two narrow rims are found
   which run parallel to each other from north to south in the western
   part of the remnant. No complex structures are seen in the interior
   region of the remnant. The X-ray spectra everywhere can be well
   represented with a power-law function of photon index ranging
   within $\Gamma =$ 2.0--2.8. A clear difference of spectra between
   the central dim region and the bright western portion is seen at
   lower energies.  This differences can be ascribed either to an
   additional thermal component with electron temperature,
   $kT_{\mathrm e}= 0.56~{\rm ~keV}$ from the center or,
   alternatively, to an increase in column density of $\Delta
   N_{\mathrm H}\sim 0.3\times 10^{22} {\rm cm}^{-2}$ in the western
   region. In the context of the recent discovery by the \emph{NANTEN}
   telescope of a molecular cloud apparently interacting with the
   western part of the supernova remnant, the second possibility seems
   to be the more likely scenario.
   
  }

   \maketitle
\section{Introduction}
The supernova remnant\, (SNR) RX\,J1713.7--3946 (also known as
G\,347.3--0.5) was discovered during the \emph{ROSAT} All-Sky survey
(Pfeffermann \& Aschenbach 1996). It is the  bright source of
nonthermal radiation in X-ray bands among the shell-type SNRs.

The \emph{ASCA} observations of Koyama et al.~(1997) showed that the X-ray
spectrum in the northwestern~(NW) rim is featureless and can be
modeled by a single power-law function with interstellar absorption.
Subsequent \emph{ASCA} observations, which covered the overall remnant
with a diameter of $\sim70\arcmin$, showed that other portions of the
remnant also have a featureless spectral shape with a complete lack of
signatures of thermal X-ray emission (Slane et al.~1999).

Based on \emph{Chandra} observations, Uchiyama et al.~(2003) resolved
the bright X-ray region in the NW rim into fine structures--a complex
network of nonthermal X-ray filaments and spots.  They showed that the
energy spectra of different parts can be represented by a power-law
with photon index which varies from site to site from $\Gamma=$
2.1--2.5, but does not correlate with the brightness distribution.
Another important result of their study was the conclusion that the
synchrotron cutoff energy is unusually high, namely it is beyond
10\,keV.

The CANGAROO collaboration reported detection 
of TeV $\gamma$-rays from the direction of the NW rim of RX J1713.7--3946
(Muraishi et al.~2000).
Later, Enomoto et al.~(2002) published results of follow up
observations by the CANGAROO 10\,m telescope.  They argue that the
obtained spectrum should be explained by $\pi^0$ decay $\gamma$-rays
rather than by Compton-upscattered $\gamma$-rays.

One of the important parameters for the interpretation of observations
is the distance to the source. Originally Koyama et al.~(1997) adopted
a distance to this SNR of 1\,kpc based on the estimation of the column
density derived from the spectral analysis of the \emph{ASCA}
data. Later, Slane et al.~(1999) argued that the source should be
located much further. It has been suggested that there exist possible
associations between the SNR RX J1713.7--3946 and adjacent molecular
clouds~(Slane et al.~1999).  Among these clouds, the so-called cloud~A
has an enhanced value of CO ($J$=2--1)/($J$=1--0) which could indicate
the physical interaction with the SNR shock~ (Slane et al.~1999; Butt
et al.~2001).  Thus, based on the kinematic distance to the cloud A,
they assumed $\sim$6\.kpc as the distance to the RX J1713.7$-$3946.
This estimate was used in most of the following studies of the
physical properties of RX~J1713.7$-$3946~(e.g. Pannuti et al.~2003,  Cassam-Chena\"{i} et al.~2004 and Lazendic et al. ~2004).

Recently,  Fukui et al.~(2003) have performed a new CO
observation at 2.6\,mm wavelength with the \emph{NANTEN} telescope
located at Las Campanas Observatory in Chile.
They discovered a cloud at a distance of 1\,kpc which surround SNR RX
J1713.7--3946.  It shows a good spatial correlation between the X-ray
and the CO emission, together with the presence of a broad CO line
down to a few arcminutes. These data favor a physical association with
RX~J1713.7--3946 rather than with the cloud at 6\,kpc.

In this paper, we report on the X-ray study of the bright rims and
interior regions of the SNR RX J1713.7--3946 by using archival data
from the \emph{XMM-Newton} satellite (P.I.\ A.~Decourchelle).  These
observations revealed the presence of a double-shell structure along
the western limb of the remnant, with enhanced absorption along this
edge (Cassam-Chena\"{i} et al. 2004). This effect naturally can be
attributed to nearby (passive or interacting with the shell) molecular
clouds.  Remarkably such a cloud has been indeed found by Fukui et
al.~(2003) using CO observations. It should be noted the
\emph{XMM-Newton} composite image presented in Fig.~2 and Fig.~3 by
Fukui et al.~ (2003) was quite similar to the image presented by
Cassa-Chena\"{i} et al.~(2004) with a clear indicaton of a
double-shell structure. However the main issue discussed in the paper
by Fukui et al.~(2003) was the correlation of the X-ray and CO peaks
on arcmin scales.

The aim of the present paper is to extend our study of the distinct
spatial and spectral features found by Uchiyama et al.~(2003) based on
the
\emph{Chandra} observations of the NW rim to the whole remnant. 

The large field-of-view~(FOV) of \emph{XMM-Newton} is a great help to
understand the nonthermal X-ray features and their three-dimensional
structures. Spatio-spectral analysis enables us to examine different
possible scenarios to investigate the spectral differences which may
arise due to, e.g. variation of the column density. Comparison
between the
\emph{XMM-Newton} image and the \emph{NANTEN} CO map in Fukui et
al.~(2003) gives a constraint on the origin of synchrotron X-ray
emission in the context of the X-ray and CO correlation.  Here, we
describe the \emph{XMM-Newton} observations in \S \ref{observation},
and present the results of our analysis in \S \ref{ana}. Finally, in \S
\ref{discussion}, we briefly discuss the implications of the results.

\section{Observation and Data reduction}\label{observation}
The observations of SNR RX J1713.7--3946 by the \emph{XMM-Newton}
satellite~(Jansen et al.~2001) were carried out in March 2001,
September 2001 and March 2002 with a total of five pointings. In this
paper, we used five data sets from the archive with the positional
identification of center\,(CE), northeast\,(NE), northwest\,(NW),
southwest\,(SW) and southeast\,(SE) as listed in Table~\ref{obs}.
These observations cover almost the whole remnant including the bright
western portion, and the central dim region. Data from the
EPIC~(European Photon Imaging Camera) which consists of two MOS
CCDs~(MOS1 \& MOS2)~(Turner et al.~2001) and the pn CCD~(Str\"uder et
al.~2001) are used for analysis.  The pixel size is 1\farcs1\, and
4\farcs1\, for the MOS and the pn CCDs, respectively, while the
mirror PSF width is 15\arcsec\, of HPD~(Half Power Diameter). All data
have been acquired with the medium filters and imaging mode, full
frame for the MOS and extended full frame for the pn CCDs.  Therefore
the temporal resolution is low, 2.5~s and 280~ms for the MOS and the
pn CCDs, respectively.

The pipeline process of event screening was carried out using Science
Analysis System~(SAS) software~(version 5.4.1) and the latest
calibration data.  In general, the data quality obtained by
\emph{XMM-Newton} depends on the condition of individual observations
because data sometimes suffer from particle flares in orbit\,(mainly
due to protons) which can not be removed by the standard pipeline
process.  These particle flares dramatically increase the count
rate by over one order of magnitude above that of quiescent
periods and cause high background in the detector.  We were careful to exclude such
intervals by using wide band~(0.5--10.0~keV) light curves
for each observation. Time periods where the count rate deviates from
the mean value during quiescent periods by $\geq2.5\ \sigma$ are removed
from subsequent analysis. Since the pn CCD is more sensitive to proton
flares than the MOS CCD, the net exposure time for the pn CCD is reduced
by a large amount~(see Table\ref{obs}).

The NE observation severely suffered from particle flares
during the observation and the net exposure time is reduced to only
2.5 ksec out of the 14 ksec of total observation time in the MOS CCDs.
Therefore, the NE observation is used only for imaging analysis
since the statistics are too poor to carry out a spectral
analysis. On the other hand, we obtain good statistics from both the
SW and CE, even with the relatively low source flux, because the
observation was carried out under good conditions in terms of the non
X-ray background~(NXB). The good quality of data enables us to perform
spatially resolved spectroscopy of the diffuse emission from dim
regions in the central part of the remnant.  Although the situation is
better than the case for the NE observation, the data from the NW
observation has a significant fraction of the NXB.  The deviation of
the count rate in the 0.5$-$10.0~keV band is 20\,\% of the mean value,
whereas it is 12\,\% in the case of SW and CE observations. We
therefore need individual estimates of the background for spectral
analysis~(see \S\ref{sec_spec}).


   \begin{table*} 
    \caption{Archival data of RX J1713.7--3946}
    \label{obs} 
    \begin{tabular}{lccc}
     \hline 
    \hline \noalign{\smallskip}
    Obs.ID & Coord~(J2000) & Exposure & Date \\
      \noalign{\smallskip}
    &  & {\scriptsize MOS1/MOS2/pn} & \\ 
    & & [ksec] & \\
     \hline
    0093670501\,(CE)&$17^{{\mathrm h}}13^{{\mathrm m}}28\fs7$,$-39^{{\mathrm d}}49^{{\mathrm m}}36^{{\mathrm s}}$&14/14/9&2/3/2001\\ 
    0093670101\,(NE)&$17^{{\mathrm h}}14^{{\mathrm m}}10\fs0$,$-39^{{\mathrm d}}26^{{\mathrm m}}00^{{\mathrm s}}$&2.5/2.6/$-^{\mathrm{a}}$&5/9/2001\\ 
    0093670201\,(NW)&$17^{{\mathrm h}}11^{{\mathrm m}}58\fs0$,$-39^{{\mathrm d}}31^{{\mathrm m}}30^{{\mathrm s}}$&14/14/10&5/9/2001\\ 
    0093670301\,(SW)&$17^{{\mathrm h}}11^{{\mathrm m}}51\fs8$,$-39^{{\mathrm d}}56^{{\mathrm m}}15^{{\mathrm s}}$&16/16/11&7/9/2001\\ 
    0093670401\,(SE)&$17^{{\mathrm h}}15^{{\mathrm m}}30\fs0$,$-40^{{\mathrm d}}00^{{\mathrm m}}57^{{\mathrm s}}$&13/13/6&14/3/2002\\
            \hline
    \end{tabular}
\begin{list}{90mm}{}
\item[$^{\mathrm{a}}$]There were little good time intervals~(GTI).
\end{list}   
   \end{table*} 
  

\section{Analysis and Results}\label{ana}
\subsection{Image analysis}\label{sec-image}
Recently, Uchiyama et al.~(2003) noted the high inhomogeneity
towards the NW portion of RX~J1713.7--3946, using the spatial resolution
of 0\farcs5 of the \emph{Chandra} satellite. The NW portion has complex, filamentary structures in the synchrotron X-ray emission,
namely, bright filaments and spots embedded in
the diffuse plateau and the dim circular void. These
structures show distinctive variations of the surface brightness on
scales down to 20\arcsec.

\begin{figure*}
   \centering
\includegraphics[angle=-0, width=10cm]{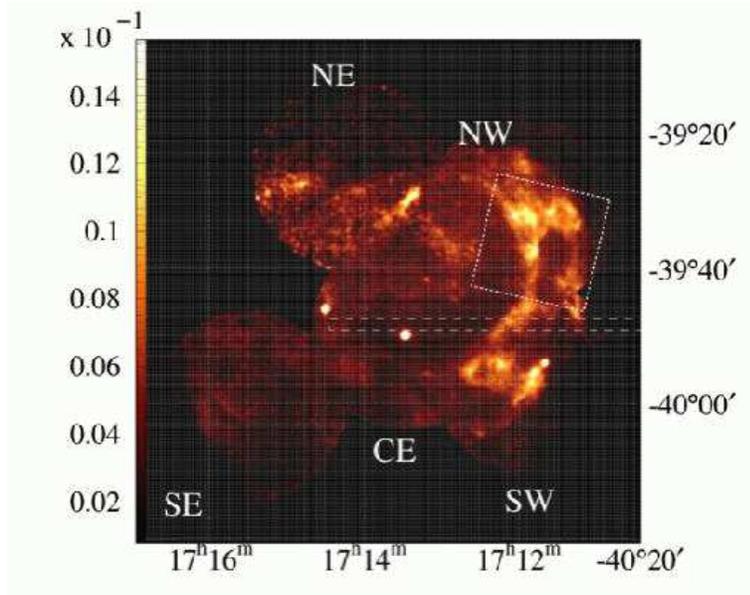}
  \caption{The mosaic image of SNR RX J1713.7--3946 using EPIC MOS1
   and MOS2 CCDs.  The 17\arcmin $\times$
   17\arcmin\, square of the \emph{Chandra} FOV is also denoted in the
   soft band image.}
   \label{image}
 \end{figure*}

Most of the entire remnant of RX~J1713.7$-$3946 is
covered with the spatial resolution of 15\arcsec\, and the large FOV of
\emph{XMM-Newton}. The spatial resolution of fifteen arcsec is sufficient to
investigate the fine structures discovered with \emph{Chandra}.
Figure~\ref{image} shows mosaic X-ray images of SNR RX~J1713.7--3946 in
0.7--2.0~keV~(soft band) 
constructed from the data obtained by the EPIC MOS1 and MOS2.  The
images are smoothed by a Gaussian with $\sigma=$\,15\arcsec\,. The
mirror vignetting and exposure time are corrected whereas the
background is not subtracted. The white square corresponds to
\emph{Chandra}\rq s 17\arcmin$\times$17\arcmin\, FOV~(Fig.~2 in
Uchiyama et al.~2003). We also construct a hard band
image~(2.0--7.0\,keV) which shows virtually identical to that in the
soft band image.

\begin{figure*}
   \centering \includegraphics[width=12cm]{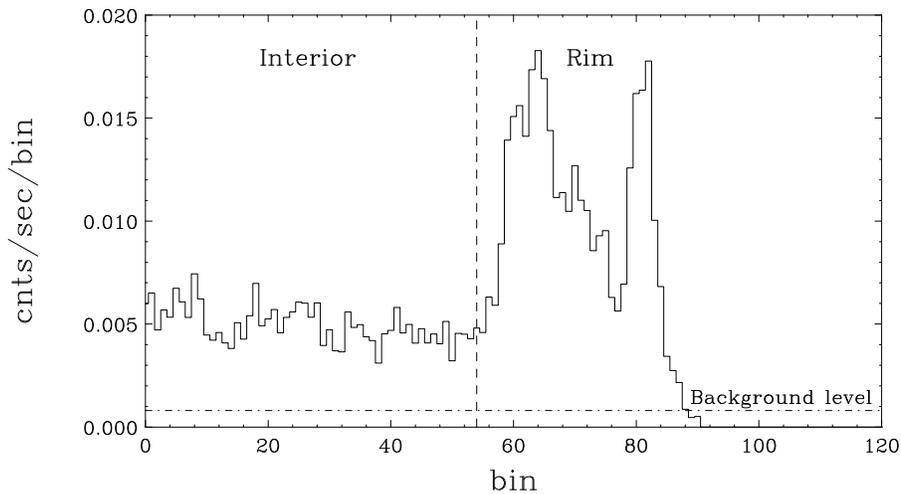}
   \caption{Radial profile of the photon counts~(0.7--2.0~keV) for the
   white-dashed rectangle region shown in Fig.~\ref{image}. It is integrated over the vertical dimension.}
 
  \label{profile}
 \end{figure*}

In the bright western portion, the thick rim with an apparent width of
$\sim$8\arcmin\, obtained with \emph{ASCA}~(Slane et al. 1999) is resolved
into distinct components~(see also Cassam-Chena\"{i} et al. 2004). Two nested narrow arc-like rims~(hereafter
we refer to them as the \lq outer rim\rq\, and \lq inner rim\rq\,) are
found. The inner rim seems to have a radius of $\sim$16\arcmin\, and is  elongated
toward the northeastern~(NE) portion. The outer rim runs parallel
to the inner rim from north to south with a gap of $\sim$8\arcmin\, whereas it is torn off in the middle west.
In the interior of the remnant, we can see diffuse,  dim X-ray
emission~(hereafter, referred to as \lq interior regions\rq\,).  In
comparison with the rim regions, no filamentary structures are seen.
Figure~\ref{profile} shows the profile of photon counts from the white-dashed 
rectangular region depicted in the soft band image of Fig.~\ref{image}, summed over the
narrow-vertical dimension of the rectangle. The interior region shows a 
quite flat profile with its deviation smaller than 20\% from the mean
value. Note that the detected counts are far from the background
level. This motivated us to perform a spatially-resolved analysis
of the spectrum with much finer angular resolution than what was
possible for the \emph{ASCA} data~(see \S~\ref{interior}).

On small scales within the rim region, complex structures are
detected in the NW portion such as bright filaments, spots and two
voids. These structures are exactly the same as discovered by
\emph{Chandra}~(Fig.6 in Uchiyama et al. 2003). Prominent filaments,
identified by 2, 3 in Fig.6 in Uchiyama et al.~(2003) are revealed to
be a part of the inner rim. In addition to the NW portion,
Fig.\ref{image} shows complex structures in the southwestern~(SW)
 portion. We find other voids in the middle west  and in the end of SW side of the remnant
enclosed by the outer rim as well as some bright filaments and
spots. There is a conspicuous plateau inserted between two
voids. The data demonstrates that the fine structures discovered in the
\emph{Chandra} FOV are a common characteristic of the SNR
RX~J1713.7--3946. These characteristic structures stand in line
between the inner and outer rims. Hereafter we refer to both the two rims and
the substructures inside them as the \lq rim region\rq\,. 



There are two point sources seen in the mosaic image. Both sources
are identified in the \emph{ROSAT} bright source catalogue, 1WGA
J1714.4--3945 associated with a star~(Pfefferman \& Aschenbach~1996)
and 1WGA 1713.4--3949. 1WGA 1713.4--3949, located near the center of
the remnant, is suggested to be the associated neutron star for the
RX~J1713.7--3946 by Lazendic et al.~(2003).

 \begin{figure*}
   \centering
\includegraphics[width=16cm]{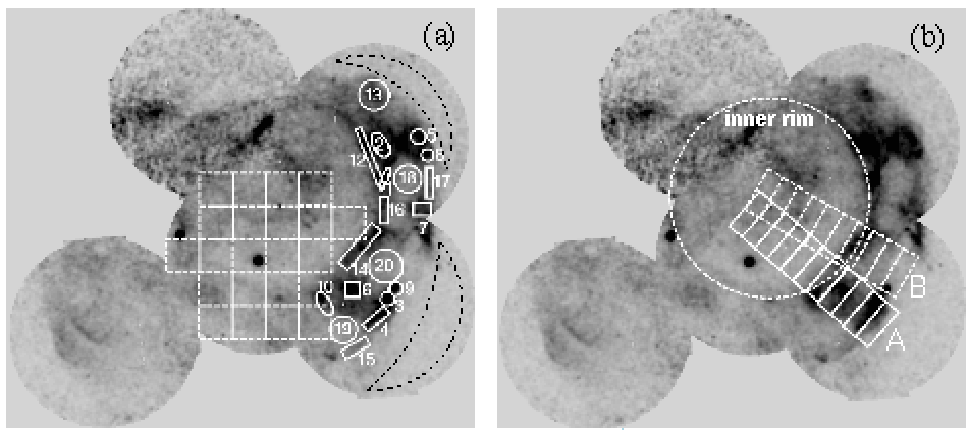}
   \caption{(a)\,The numbered regions are selected for the spectral
  analysis of the rim region and the dashed squares are selected for the
  spectral analysis of the interior region. The dashed black curves
  denote the background region for NW and SW observations, respectively. (b)\,
   A and B are two sets of 10 rectangle regions selected for examination
  of the possible spectral variation in the radial direction. The named regions of NWb, NWd, SWb and SWd are selected to estimate absolute fluxes~(\S\ref{sec-radial}). The apparent   circular shape of the inner rim is also depicted.}
 \label{map} 
\end{figure*}

\subsection{Spectral analysis}\label{sec_spec}
As suggested by previous observations, the entire region of
RX~J1713.7--3946 is dominated by nonthermal X-ray emission with a
complete lack of thermal emission~(Koyama et al.~1997;\, Slane et
al.~1999). Uchiyama et al.~(2003) demonstrated that the X-ray spectra
are best fit by hard power-law spectra with a photon index ranging
2.1--2.5 which does not correlate with significant brightness
variations within the NW rim inside the \emph{Chandra} FOV.  They
argued, furthermore, that the observed hard power-law seems to require
the synchrotron cutoff energy to be above 10~keV which needs a high
shock speed exceeding 5\,000\,km\,${\rm s}^{-1}$.

In the spectral analysis, especially for extended sources, a careful
estimation of the background~(BG) is necessary.  Since the SNR
RX~J1713.7--3946 is located near the galactic plane, we need to
extract the BG data from a source free region in the same FOV.  In
the analysis, the BG data is extracted from the individual FOV denoted
by the dashed curve both in NW and SW~(Fig.~\ref{map}~for the MOS1 and
MOS2 data) because the signal to noise ratio is different between the
NW observation and SW observation.  These data include NXB as well as
cosmic X-ray background~(CXB) and the galactic diffuse emission.  For
the pn CCD, the BG data are extracted from the outside of SNR RX
J1713.7--3946 which is in the outer part of the detector. We have
examined the BG flux variation in the detector plane using archived
Lockman Hole data. We find no significant variation for the continuum
over the 5~keV\, in which the NXB is dominant. The intensity of
fluorescent lines of the NXB, especially 7.0--9.0~keV\,, in the pn CCD
has a significant position dependency in the detector~(e.g. Lumb et
al.~2002). No significant source fluxes are detected above 9.0~keV\,. We,
therefore, ignore the data above 7.0~keV in the subsequent spectral
analysis. In the low energy part~($\leq$\,1 keV), the NXB is relatively
constant~($\pm 20 \%$) and a high signal to noise ratio above 0.7
keV\, is obtained. In order to derive the spectral parameters, we
perform simultaneous spectral fitting with all of the EPIC instruments
from 0.7 keV\, up to the 7.0 keV\, intervals.  For the CE observation,
the same BG data as for the  SW observation were used since the data have no
source free regions and the light curve shows a similar level with that
of the SW observation~(see \S \ref{observation}).


\subsubsection{Rim regions}
We select 20 characteristic regions in the rim region for the study of  spectral variations~(Fig.~\ref{map}~(a)). From all spectra, we
detect only featureless continua without any line emission. Spectra
are fitted by a power-law function with photoelectric absorption
along the line-of-sight, $N_{\mathrm H}$.  
\emph{XMM-Newton} has a large effective area, much
larger than that of \emph{Chandra}, especially in the low energy
part~($\leq 1 {\rm keV}$). This enables us to determine $N_{\mathrm
H}$ more precisely, because $N_{\mathrm H}$ is strongly constrained by
spectra below 1~keV. On the other hand, RX~J1713.7--3946 is located in
the inhomogeneous region of the galactic plane and is suggested to
have a physical interaction with the molecular cloud. It is important
to examine the possible variation of $N_{\mathrm H}$ as well as
$\Gamma$. We therefore set both photon index, $\Gamma$\, and
absorption column density, $N_{\mathrm H}$ free in the spectral
fitting.

Spectra from representative region \lq 6\rq\, are shown in
Fig.~\ref{spectra}~(a) with the best fit model as a typical example.
We show the spectral parameters obtained for these regions depicted
with filled circles in Fig.~\ref{rim_para}.  It is demonstrated that
all spectra are fitted by a power-law function with photon index
$\Gamma$\, ranging from 2.1--2.6\,, with a somewhat large
$\chi^2_{\nu}\sim1.38$ for $\nu\le140$.  Best-fit values of the
absorbing column density, $N_{\mathrm H}$, are similar in almost all
cases, with average values of $N_{\mathrm H}
\sim 0.78 \times 10^{22}~{\rm cm^{-2}}$. These values are consistent
with the \emph{ASCA} results obtained from the spectra integrated over the NW
and SW rims.~(Koyama et al.~1997;\,Slane et al.~1999).

Based on the \emph{Chandra} observations, Lazendic et al.~(2004)
reported that the spectral shape in the brighter regions is flatter
than those in the fainter regions (Fig.~B3 in Lazendic et al.~2004)
adopting a fixed value of column density.  In order to verify the
significance of the spectral variation both for $\Gamma$ and for
$N_{\mathrm H}$, we tried to fit spectra with a fixed $N_{\mathrm H}$
of $ 0.78 \times 10^{22} {\rm cm^{-2}}$, an average value for the rim
regions. Although similar results are obtained in almost of all cases
as denoted with circles in the top and bottom panels of
Fig.~\ref{rim_para}, we find that $\chi^2_{\nu}$ increase remarkably
for several regions such as regions ``4'' and ``15''  in which
there could be true variation of  column density in the small scale of $\sim$arcmin.  By treating
$N_{\mathrm H}$ as a free parameter  we obtain significantly large
$N_{\mathrm H}$, greater than 1.0$\times\, 10^{22}~{\rm cm^{-2}}$ for
these regions  which possibly corresponds to the region where the
maximums of CO emission was reported~(Fukui et al.~2003).


%

\subsubsection{Interior region}\label{interior}
\begin{figure*} \centering
 \includegraphics[angle=0,width=17cm]{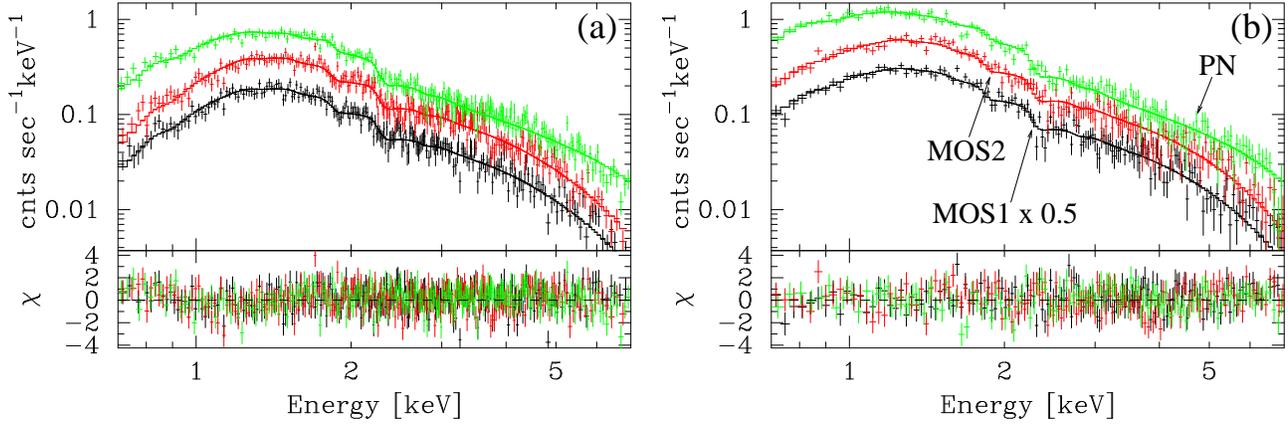}
  \caption{ Representative spectra obtained by all CCD detectors, MOS1~(black), MOS2~(red) and pn~(green) are shown as characteristic examples both for the rim region~(a) and the interior region~(b). In both (a) and (b), MOS1 spectra are scaled to 0.5 times for clear look.}
  \label{spectra}
      \end{figure*}

\begin{figure*}
   \centering
   \includegraphics[angle=0,width=15cm]{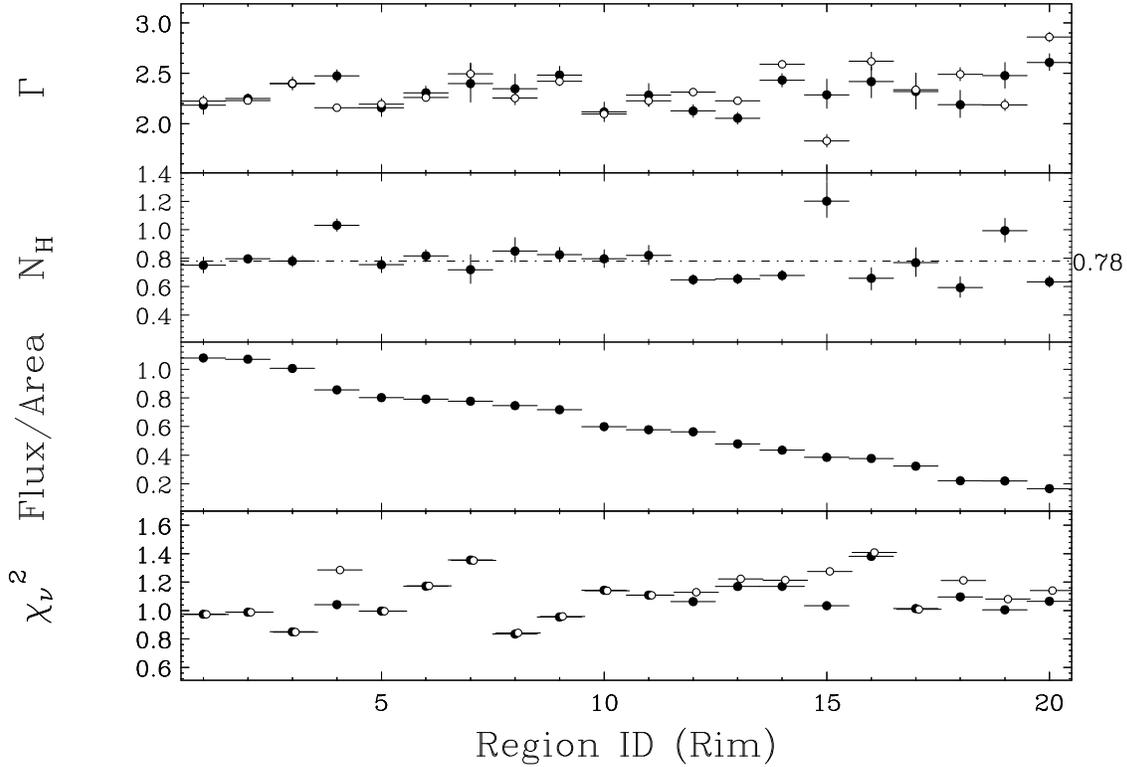}
   \caption{Results of spectral fits with an absorbed power-law model
   for selected regions in the western bright portion depicted by filled circles. From top to bottom panels, plotted are the best-fit values~(with their 90\% errors) of
   photon index, $\Gamma$, and absorbing column density, $N_{\mathrm H}$ in
   units of $10^{22}\,{\rm cm}^{-2}$, unabsorbed flux in 1.0--10~keV band
   in units of $10^{-12} {\rm erg\,cm^{-2}\,s^{-1}\,arcmin^{-2}}$ and $\chi^2_{\nu}$.  Resultant $\Gamma$ and $\chi^2_{\nu}$ adopting by the fixed $N_{\mathrm H}$ of 0.78\,$\times\,10^{22}{\rm cm^{-2}}$ are shown by circles in the top and bottom panels, respectively.}
  
   \label{rim_para}
	\end{figure*}

For finding out the spectral property in the interior region, we select
an annular region from 7\farcm5\, to 10\arcmin\, intervals around the
central point source and perform the spectral fitting with power-law and
photoelectric absorption models. Fig.~\ref{spectra}~(b) shows the
obtained spectra which show featureless continua without line emission 
for the best-fit models. We find that the spectral shape is different
from that in Fig.~\ref{spectra}~(a)~(rim region) in the low energy
part~($\leq$ 1\,keV). The resultant photon index is $\Gamma\,= 2.21 \pm
0.05$ and the absorption column density is  $N_{\mathrm H} = 0.46 \pm 0.02
\times 10^{22}~{\rm cm^{-2}}$.  In order to examine possible
spectral variations in the interior region, we  divide the region
into 20 squares of 5\arcmin\,$\times$ 5\arcmin\,~(Fig.~\ref{map}~(a)) and
perform a systematic spectral fitting. The resultant photon indices,
$\Gamma$, vary ranging from 2.0--2.8 from place to place. It seems to have no
correlation with either absorption column density, $N_{\mathrm H}$, or
flux.

\begin{figure*}
\centering 
\includegraphics[angle=0,width=14cm]{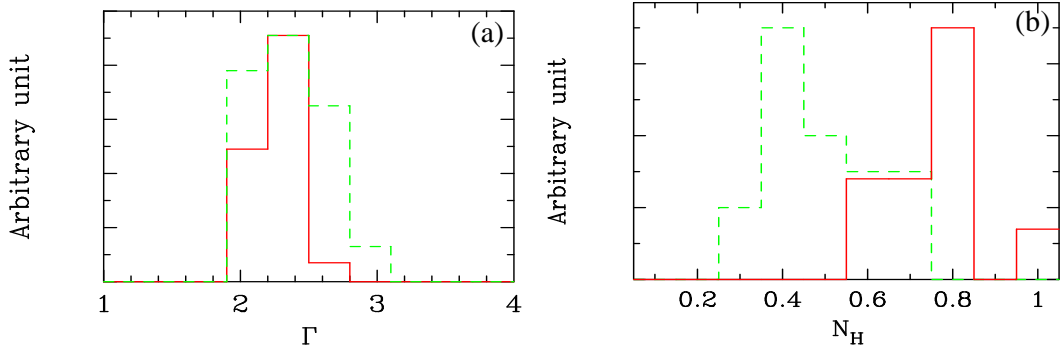} 
\caption{Best-fit parameter distribution of $\Gamma$~(left) and $N_{\rm
H}$~(right) both for the rim region~(solid line) and the interior
region~(dashed line).}

\label{hist}
\end{figure*}

In Fig.~\ref{hist}, we show the distributions of parameters $\Gamma$
and $N_{\mathrm H}$ obtained with the best-fit model. It is clear that
each region has a similar distribution of $\Gamma$ with an average of
$\Gamma \sim 2.4$. However $N_{\mathrm H}$ shows obvious discrepancies
with average of $N_{\mathrm H} = 0.46\times\,10^{22}~{\rm cm^{-2}}$ for
the interior region and $N_{\mathrm H} = 0.78\times\,10^{22}~{\rm
cm^{-2}}$ for the rim region.  The increase of column density, $\Delta
N_{\mathrm H} \sim\,0.32\times\,10^{22}~{\rm cm^{-2}}$ implies that the
rim region is covered by an additional medium which absorbs X-ray
emission, such as a dense molecular cloud.

\begin{figure*} \centering
  \includegraphics[angle=-90,width=8cm]{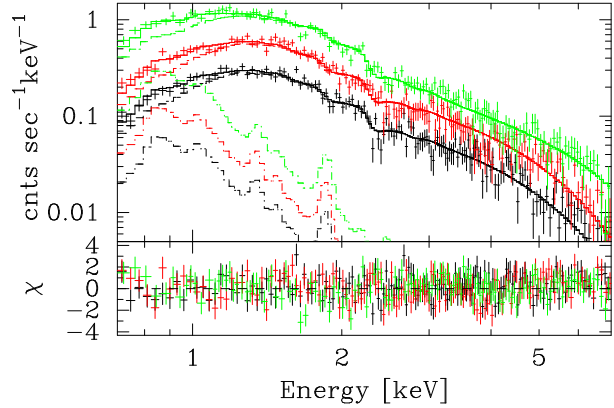}
  \caption{Spectrum of the interior region~(same data as shown in
  Fig.~\ref{spectra}~(b)) with the best-fit model containing a thermal
  component in addition to the absorbed power-law model.}

  \label{thermal}
    \end{figure*}

One of the unique features of RX~J1713.7--3946 is the complete lack of
 measurable thermal emission which is expected from the hot gas that
resides in SNRs. Nevertheless one may assume that the observed spectral
differences from the rim and interior region at  low energies
part can be caused by an additional thermal component from the center
of the remnant due to a limited energy resolution of CCDs. Assuming that
the interstellar absorption of $\sim 0.78\times\,10^{22}\,{\rm
cm^{-2}}$~(the average value of $N_{\mathrm H}$ at the rim region) is constant
 over the entire remnant, we can calculate the largest possible amount of the
thermal component. We perform a spectral fit for the same spectra as
shown in Fig.~\ref{spectra}~(b) with an additional thermal component of a MEKAL model~(XSPEC v11.0). We derive the best-fit values for different
$N_{\mathrm H}$, e.g. 0.76\,$\times\,10^{22}~{\rm cm^{-2}}$,
$0.78\,\times\,10^{22}~{\rm cm^{-2}}$ and $0.80\,\times\,10^{22}~{\rm
cm^{-2}}$. We set the photon index, $\Gamma$, as a free parameter and fix
the abundances to solar values. Statistically acceptable
results~($\chi^2_{\nu} \sim\,1.1$) are obtained from the
fitting. The best-fit model in the case of $N_{\mathrm H} =
0.76\,\times\,10^{22}\,{\rm cm^{-2}}$ is shown in
Fig.~\ref{thermal}. The resultant parameters are $kT_{\rm e}=$
0.56--0.57 keV, $\Gamma =$ 2.37--2.39 for each case of $N_{\mathrm
H}$. The emission measure, EM $=\,n_{\mathrm e} n_{\mathrm H} V/4\pi d^2$ is
calculated to be $\sim~3.2 \times\,10^{11}~{\rm cm^{-5}}$ where $n_{\rm
e}$ and $n_{\mathrm H}$ are the mean number density of electron and
hydrogen, $V$ is the volume and $d$ is the distance to the remnant.
Assuming $n_{\mathrm e}$ is 1.2 times  $n_{\mathrm H}$ and a uniform density
distribution inside the sphere of 16\arcmin\, in radius~(inner rim), we
derive $n_{\mathrm e} \sim 0.1\,d_{1}^{-1/2} {\rm cm^{-3}}$, where $d_{1}$
is the distance normalized to 1~kpc.

We should note, however, the hypothesis of the additional thermal
component, which requires a very large interstellar absorption column
density, does not match with the X-ray spectrum of the central point
source (1WGA 1713.4--3949). It is well fitted by a blackbody model
with  photoelectric absorption. The best-fit value~($\chi^{2}_{\nu} = 1.14$) of $kT_{\mathrm {e}}$ is 0.40 $\pm$ 0.01\,keV\, and that of
$N_{\mathrm H}$ is 0.37 $\pm$ 0.03\,$\times\,10^{22}\,{\rm
cm^{-2}}$. The resultant absorption column density, $N_{\mathrm H}$, is
in rather good agreement with that found towards the interior
region. Even so, with this approach, we can derive an important upper
limit on the flux of the thermal component. We note that the possible
thermal emission is very subtle, such as $0.4\,\times\,10^{-12}\,{\rm
erg\,cm^{-2}\,sec^{-1}\,arcmin^{-2}}$ of unabsorbed flux in
0.6--10.0\,keV\, X-ray band per solid angle.  This is only 10\,\% of the
nonthermal emission. It is quite different from  the case of SN~1006. The
X-ray emission of the central region of SN~1006 is completely due to
thermal emission with no nonthermal component. Even near the rim,
the thermal emission is observed to be comparable to that of the nonthermal emission of
$\mbox{0.7--0.8}\,\times\,10^{-12}\,{\rm
erg\,cm^{-2}\,sec^{-1}\,arcmin^{-2}}$ in the 0.6--10.0\,keV\, band~(Bamba et al.~2003).

\subsection{Radial profile of the spectral variation}\label{sec-radial}
\emph{XMM-Newton} observations enabled us to quantify the spectral
variation in the dim interior region of RX~J1713.7$-$3946.  We observe
that the spectrum in the interior region is dominated by nonthermal
X-ray emission as pointed out by \emph{ASCA} observations~(Koyama et
al.~1997;\,Slane et al.~1999). This is very unusual because thermal
emission from a hot plasma is generally expected as seen in SN~1006.  At
least in the framework of the standard diffusive shock acceleration
model, in which the electrons accelerated in the shell hardly could
achieve, due to severe radiative losses, the central parts of the
remnant. we should expect strong steepening of the synchrotron
spectrum moving towards the center but it is not the case for
RX~J1713.7$-$3946. It is therefore important to explore the
three dimensional structure of the remnant, in particular to examine
whether the emission from the interior region could be the result of
geometrical effects, i.e. the projection of the shell, or intrinsic
origin due to relativistic electrons interacting with the ambient
magnetic fields.

In order to investigate the radial variation of the nonthermal X-ray
emission, we extract spectra from two sets of 10 rectangle regions with
2\farcm5 intervals~(Fig.~\ref{map}(b)). (A) is selected such that the
regions run across both the inner and outer rims, whereas (B) is
selected to run across only the inner rim.  Spectral fits are performed
using the same model of the previous analysis and consistent results
for best-fit parameters of $\Gamma$ and $N_{\mathrm H}$ are obtained.

\begin{figure*} \centering
\includegraphics[angle=0,width=15cm]{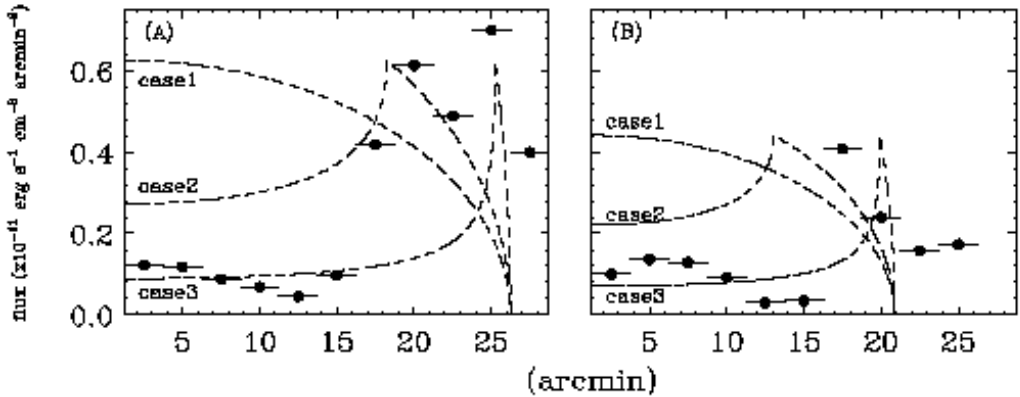}
\caption {Radial profiles for two sets of 10 rectangle regions, (A) and
(B)~( depicted in Fig.~\ref{map}(b)). The dashed curves are three cases to
examine a possible shell thickness assuming uniform distribution of
the volume emissivity.}
  \label{projection}
\end{figure*}

In Fig.~\ref{projection}, we construct radial profiles of the surface
brightness. We examine three simple cases assuming uniform
distribution of the volume emissivity;\,the sphere~(case 1), the
8\arcmin-thick shell~(case 2) and 1\arcmin-thick shell~(case 3). The
one arcmin of shell thickness roughly corresponds to the width of the
inner or outer rims whereas the eight-arcmin of shell thickness
corresponds to that of the bright western portion containing both the
inner and outer rims.  None of our assumptions can explain the
observed profiles of the surface brightness;  there is a broad shell and
the interior is about one order of magnitude dimmer than the rim.  Although there is
little doubt that the geometrical effect of the shell brighten
contributes non-negligibly to the flux observed from the dim interior
region, it is difficult to obtain a quantitative estimate of the ratio
of the shell emission to the other contribution given the complex
shape of the remnant.  Moreover, we note that the bright compact
features appear only in the rim region.  We may conclude that the surface
brightness variation is somewhat contributed to by the intrinsic origin
and/or difference of the volume emissivity.


In Table~\ref{flux}, we listed the best-fit values and the absolute fluxes which is
accurate to within about 10\% in the current investigation of on-board calibration\footnote{detailed in
http:/xmm.vilspa.esa.es/docs/documents/XMM-SOC-CAL-TN-0018}. In the
rim regions, the resultant fluxes are integrated for characteristic
regions such as NWb, NWd, SWb and SWd denoted in Fig.~\ref{map}(b).  This
is consistent with the \emph{ASCA} results~(Uchiyama et al. 2002).  In order to
estimate how much emission totally comes from the interior dim region, we
sum up the resultant fluxes for  twenty  square regions of 5\arcmin\,$\times$\,5\arcmin~(see
\S\ref{interior}).  

%
%

   \begin{table*} 
    \caption{Summary of the luminosity from representative regions in RX J1713.7--3946}
    \label{flux} 
    \begin{tabular}{cccccc}
     \hline 
    \hline \noalign{\smallskip}
    Region & area  &$\Gamma$&$N_{\rm H}$& $F_{{\rm 1-10\,keV}}$ & $L_{X}$\\
      \noalign{\smallskip}
    &(arcmin$^2$)&&$(10^{22} {\rm cm}^{-2})$&${\rm (erg\,s^{-1}\,cm^{-2})}$ & ${\rm (erg\,s^{-1})}$ \\ 
     \hline
    NWb & 125 & 2.28$\pm0.02$&0.72$\pm0.01$& 7.3$\times 10^{-11}$ & 8.7$\times10^{33}d_{1}^{2}$\\
    NWd & 64 & 2.12$\pm0.04$&0.68$\pm0.02$& 2.8$\times 10^{-11}$ & 3.4$\times10^{33}d_{1}^{2}$\\
    SWb & 125 & 2.38$\pm0.03$&0.83$\pm0.01$& 6.2$\times 10^{-11}$& 7.4$\times10^{33}d_{1}^{2}$\\
    SWd & 77 & 2.54$\pm0.04$&0.68$\pm0.02$& 1.9$\times 10^{-11}$ & 2.3$\times10^{33}d_{1}^{2}$\\
    Interior & 500 & 2.39$^{\mathrm {a}}$&0.46$^{\mathrm {a}}$& 5.4$\times 10^{-11}$& 6.5$\times10^{33}d_{1}^{2}$\\
            \hline
    \end{tabular}
\begin{list}{90mm}{}
\item[Note.] $F_{{\rm 1-10\,keV}}$ is the unabsorbed flux in the 1--10\,keV band. $d_{1}$ is the distance to the source normalized to 1\,kpc. 
\item[$^{\mathrm{a}}$] average values with totally 20 squires of 5\arcmin\,$\times\,$5\arcmin.
\end{list}   

   \end{table*}

\section{Discussion}\label{discussion}
We examined the SNR RX~J1713.7--3946 by using the \emph{XMM-Newton}
data to further investigate the remarkable features in nonthermal X-ray
emission discovered by Uchiyama et al.\ (2003).
The \emph{XMM-Newton} X-ray image demonstrates complex morphological
structures such as filaments and voids in the bright western region,
while relatively weak and smooth emission around the center of the
remnant.  The X-ray spectra everywhere in the remnant are successfully
represented by a power-law with a photon index ranging from 2.0--2.8. We
found no line features in the X-ray spectrum anywhere.  The low energy part
($\le$\,1\,keV) of the X-ray spectra in the dim interior region
systematically differs from that in the western bright portion, which is
attributable either to an additional \emph{absorption} in the western
spectra or to an additional \emph{thermal} component in the interior
region.
The surface brightness of nonthermal X-ray emission of RX~J1713.7$-$3946
has distinctive variations across the remnant both on large scales
(Pfeffermann \& Aschenbach~1996; Slane et al.~1999), showing broad
shell-like morphology, and on small scales down to 20\arcsec\,
(Uchiyama et al.~2003).  There is no correlation between photon
indices, absorption column density and surface brightness in the
western bright~(rim) region. A similar spectral shape of power-law
suggests that the same emission mechanism --- presumably synchrotron
radiation given the observed luminosity and spectral shape --- is
operating everywhere.

The observed surface brightness variations can be ascribed to different
volume emissivities from place to place, or to different path lengths
along the line-of-sight with similar volume emissivity. 
Dimmer parts of
the remnant have a surface brightness typically an order of magnitude
lower than the bright plateau.  Adopting a typical depth of the bright
plateau of 10\arcmin\, inferred from an apparent thickness, the dimmer
portions should have a thin shell-like structure of a thickness of
1\arcmin\, if we assume the same volume emissivity.  However, because we
do not find such a thin shell at the periphery of the remnant, it seems
difficult to regard all of the dim regions as a thin bright shell.  Instead
of the effect of different path length, we therefore prefer to invoke
spatial variation of volume emissivity; the western part would be
bright thanks to an enhanced emissivity.  Moreover, the presence of
spots indicates significant enhancement of the volume emissivity of
nonthermal X-rays.

The high X-ray volume emissivity can be caused by concentration of
relativistic electrons and/or large magnetic fields.  The luminosity
of synchrotron X-ray emission most likely reaches its maximum value
given by the injection rate of electrons at least on large
scales. Therefore the enhanced X-rays in the bright western portion
would not be due solely to large magnetic fields, but result primarily
from a higher electron injection rate there.

\begin{figure*} 
\centering 
\includegraphics[angle=0,width=8cm]{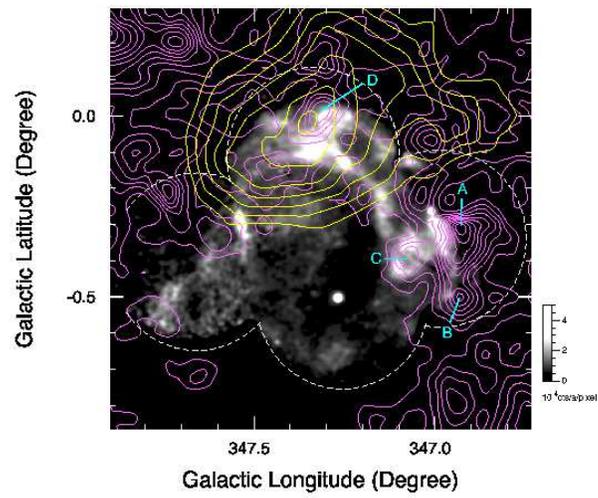} 
\caption{The same figure as Fig.2 of Fukui et al.~(2003). The CO distribution at a velocity range from $-11$ to $-3{\rm km s^{-1}}$ is depicted by the pink contour. TeV $\gamma$-ray significance map is  superposed as the  yellow contour. The X-ray image is also shown; it is the soft band image obtained by XMM-Newton as shown in Fig.1.}
\label{nanten}
\end{figure*}

Fukui et al.~ (2003) have recently reported the discovery of a molecular
cloud at 1\,kpc interacting with the SNR RX~J1713.7--3946, based on CO
observations made with the \emph{NANTEN} telescope.
The observations (at 2.6 mm wavelength) covered the entire SNR region
with 2\arcmin\, grids achieving a high sensitivity that enabled them
to detect a column density down to $1.7\times 10^{20}\, \rm atoms\
cm^{-2}$.  A morphological correspondence between the newly-discovered
molecular cloud of velocity range from $-11$ to $-3\, {\rm
km\,s^{-1}}$ and X-ray emission is evident; a ``hole'' in the CO
distribution matches the overall X-ray morphology (see Fig.1 in Fukui
et al.~2003).  In Fig.~\ref{nanten}~(the same with Fig.~2 of Fukui et
al.~ 2003), a close-up view of the CO map is compared with the X-ray
image taken by \emph{XMM-Newton} which referred to this study.  A
remarkable correspondence can be seen on an arcmin scale,
particularly in the western portion.

We found possible evidence that the X-ray spectra in the western
bright  part suffer from larger absorption column densities, $\Delta N_{\mathrm H}\,\simeq\,$0.3--0.4$\,\times\, 10^{22}\,\rm cm^{-2}$, than
the interior region~(see \S\ref{interior}).  The typical intensity of
the CO line of $W_{\rm CO} \sim 8\ \rm K\ km\ s^{-1}$ detected by
\emph{NANTEN} toward the X-ray bright part can be used to an atomic hydrogen column density of $N_{\mathrm H}\,\sim\,0.3\,\times\,10^{22}\
\rm cm^{-2}$ with the conversion relation from the CO line intensity to
the molecular hydrogen column density of $N_{\mathrm H_2} =
2\,\times\,10^{20}\, (W_{\rm CO}/\rm K\ km\ s^{-1})\, \rm cm^{-2}$ (Bertsch
et al.~1993), which nicely agrees with the inferred additional
absorption column density toward the western bright portion.  With the
current data, we can also explain the change of the low energy part of
X-ray spectra by introducing an additional thermal component from
inside the remnant.  We need further observations to disentangle the
two possibilities.

The presence of the molecular cloud being coincident with the X-ray bright
regions also would account for the enhanced nonthermal emissivity in the
western part.  The striking association between the CO peaks (A--D) and
X-ray bright features (it is important to note that both CO and X-ray
emissions have been detected with superior statistical significance)
suggests that the high X-ray emissivities would be caused by enhanced
electron acceleration and/or compressed magnetic fields at secondary
shocks resulting from the collisions of the SNR blast wave and the
molecular cloud.  Yet another possibility, although quite unusual, is
that the nonthermal X-ray emission emerges directly from the molecular
cloud.  With this interpretation, the high X-ray emissivities would
originate in the molecular cloud itself.  We will discuss the origin of
the nonthermal X-ray emission in RX~J1713.7--3946 in a forthcoming paper.

\section{Conclusion} 
\emph{XMM-Newton} observations show the existence of fine structures
throughout the entire remnant, especially in its western portion in which interaction with a dense molecular cloud is strongly suggested  the recent \emph{NANTEN} observations.  
Spectra everywhere in the remnant can be represented by power-law
functions with photon index ranging from $\Gamma =$ 2.0--2.8 assuming some
absorption in line-of-sight of $10^{21}$--$10^{22}\,{\rm cm^{-2}}$. We
found that there is no correlation between photon indices and the
brightness variation.
It is found that the $N_{\mathrm H}$ obtained for the rim regions is systematically larger than that obtained for the interior regions. The differences of spectral shape between the rim and
the interior region at lower energies can be explained by an
additional column density, $\Delta N_{\mathrm H}\,\simeq\,$0.3--0.4\,$\times\,10^{22}\,\rm cm^{-2}$, for the rim region. This is in good agreement with the
molecular cloud recently discovered by the \emph{NANTEN} telescope. The
spectral differences can be also described by introducing an additional
thermal emission from  the interior region. Although we treat this as
a less likely scenario, given the strong evidence of interaction of a
dense molecular cloud with the shell of the supernova remnant, it
gives a quite robust upper limit on the flux of the thermal X-ray
emission component from this supernova remnant.  The thermal X-ray
flux appears to be an order of magnitude smaller than the nonthermal flux of
RX~J1713.7--3946.


\end{document}